\def\answ{b }
\input harvmac
\input labeldefs.tmp
\writedefs
\overfullrule=0pt

\input epsf
\def\fig#1#2#3{
\xdef#1{\the\figno}
\writedef{#1\leftbracket \the\figno}
\nobreak
\par\begingroup\parindent=0pt\leftskip=1cm\rightskip=1cm\parindent=0pt
\baselineskip=11pt
\midinsert
\centerline{#3}
\vskip 12pt
{\bf Fig. \the\figno:} #2\par
\endinsert\endgroup\par
\goodbreak
\global\advance\figno by1
}
\newwrite\tfile\global\newcount\tabno \global\tabno=1
\def\tab#1#2#3{
\xdef#1{\the\tabno}
\writedef{#1\leftbracket \the\tabno}
\nobreak
\par\begingroup\parindent=0pt\leftskip=1cm\rightskip=1cm\parindent=0pt
\baselineskip=11pt
\midinsert
\centerline{#3}
\vskip 12pt
{\bf Tab. \the\tabno:} #2\par
\endinsert\endgroup\par
\goodbreak
\global\advance\tabno by1
}
\def\der{\partial}
\def\d{{\rm d}}

\def\E#1{{\rm e}^{\textstyle #1}}%

%

\def\Gam{{\mit\Gamma}}

\def\HS{Hubbard--Stratonovitch}
\def\Th{Thistlethwaite}
%
\def\pre#1{ (preprint {\tt #1})}
%
%
%
\lref\HTW{J. Hoste, M. Thistlethwaite and J. Weeks, 
{\sl The First 1,701,936 Knots}, {\it The Mathematical Intelligencer}
20 (1998) 33--48.}
\lref\AV{I.Ya. Arefeva and I.V. Volovich, 
{\sl Knots and Matrix Models}, {\it Infinite Dim.
Anal. Quantum Prob.} 1 (1998) 1\pre{hep-th/9706146}).}
\lref\BIPZ{E. Br{\'e}zin, C. Itzykson, G. Parisi and J.-B. Zuber, 
{\sl Planar Diagrams}, {\it Commun. Math. Phys.} 59 (1978) 35--51.}
\lref\BIZ{D. Bessis, C. Itzykson and J.-B. Zuber, 
{\sl Quantum Field Theory Techniques in Graphical Enumeration},
{\it Adv. Appl. Math.} 1 (1980) 109--157.}
\lref\DFGZJ{P. Di Francesco, P. Ginsparg and J. Zinn-Justin, 
{\sl 2D Gravity and Random Matrices}, {\it Phys. Rep.} 254 (1995)
1--133.}
\lref\tH{G. 't Hooft, 
{\sl A Planar Diagram Theory for Strong 
Interactions}, {\it Nucl. Phys.} B 72 (1974) 461--473.}
\lref\HTW{J. Hoste, M. Thistlethwaite and J. Weeks, 
{\sl The First 1,701,936 Knots}, {\it The Mathematical Intelligencer}
20 (1998) 33--48.}
\lref\MTh{W.W. Menasco and M.B. \Th, 
{\sl The Tait Flyping Conjecture}, {\it Bull. Amer. Math. Soc.} 25
(1991) 403--412; 
{\sl The Classification of Alternating 
Links}, {\it Ann. Math.} 138 (1993) 113--171.}
\lref\Ro{D. Rolfsen, {\sl Knots and Links}, Publish or Perish, Berkeley 1976.}
\lref\STh{C. Sundberg and M. Thistlethwaite, 
{\sl The rate of Growth of the Number of Prime Alternating Links and 
Tangles}, {\it Pac. J. Math.} 182 (1998) 329--358.}
\lref\Tutte{W.T. Tutte, {\sl A Census of Planar Maps}, 
{\it Can. J. Math.} 15 (1963) 249--271.}
\lref\Zv{A. Zvonkin, {\sl Matrix Integrals and Map Enumeration: An Accessible
Introduction},
{\it Math. Comp. Modelling} 26 (1997) 281--304.}
\lref\KM{V.A.~Kazakov and A.A.~Migdal, {\sl Recent progress in the
theory of non-critical strings}, {\it Nucl. Phys.} B 311 (1988)
171--190.}
\lref\KP{V.A.~Kazakov and P.~Zinn-Justin, {\sl Two-Matrix Model with
$ABAB$ Interaction}, {\it Nucl. Phys.} B 546 (1999) 647
\pre{hep-th/9808043}.}
\lref\ZJZ{P.~Zinn-Justin and J.-B.~Zuber, {\sl Matrix Integrals
and the Counting of Tangles and Links},
to appear in the proceedings of the 11th 
International Conference on Formal Power Series and Algebraic 
Combinatorics, Barcelona June 1999
\pre{math-ph/9904019}.}
\lref\ZJZb{P.~Zinn-Justin and J.-B.~Zuber, {\sl On the Counting of Colored
Tangles}, {\it Journal of Knot Theory and its Ramifications} 
9 (2000) 1127--1141\pre{math-ph/0002020}.}
\lref\PZJ{P.~Zinn-Justin, {\sl Some Matrix Integrals
related to Knots and Links}, proceedings
of the 1999 semester of the MSRI ``Random Matrices
and their Applications'', MSRI Publications Vol. 40 (2001)
\pre{\tt math-ph/9910010}.}
\lref\PZJb{P.~Zinn-Justin, {\sl The Six-Vertex Model on
Random Lattices}, {\it Europhys. Lett.} 50 (2000) 15--21\pre{cond-mat/9909250}.}
\lref\IK{I.~Kostov, {\sl Exact solution of the Six-Vertex
Model on a Random Lattice}, {\it Nucl. Phys.} B 575 (2000) 
513--534\pre{hep-th/9911023}.}
\lref\KAUF{L.H.~Kauffman, {\sl Knots and Physics},
World Scientific Pub Co (1994).}
\lref\KAUFb{L.H.~Kauffman, {\sl Virtual Knot Theory}\pre{math.GT/9811028}.}
\lref\JZJ{J.~L.~Jacobsen and P.~Zinn-Justin, {\sl A Transfer Matrix approach to the Enumeration of Knots}\pre{math-ph/0102015};
{\sl A Transfer Matrix approach to the Enumeration of Colored Links}\pre{math-ph/0104009}.}
\lref\EK{B.~Eynard and C.~Kristjansen, {\sl More on the exact solution of the
$O(n)$ model on a random lattice and an investigation of the case $|n|>2$},
{\it Nucl. Phys.} B 466 (1996) 463--487\pre{hep-th/9512052}.}
\lref\KoS{I.K.~Kostov, {\it Mod. Phys. Lett.} A4 (1989), 217\semi
M.~Gaudin and I.K.~Kostov, {\it Phys. Lett.} B220 (1989), 200\semi
I.K.~Kostov and M.~Staudacher, {\it Nucl. Phys.} B384 (1992), 459.}
\lref\KPZ{V.~G.~Knizhnik, A.~M.~Polyakov and A.~B.~Zamolodchikov,
{\sl Fractal structure of 2D quantum gravity},
{\it Mod.~Phys.~Lett.~A} 3, 819--826 (1988);
F.~David,
{\sl Conformal field theories coupled to 2D gravity in the conformal gauge},
{\it Mod.~Phys.~Lett.~A} 3, 1651--1656 (1988);
J.~Distler and H.~Kawai,
{\sl Conformal field theory and 2D quantum gravity},
{\it Nucl.~Phys.} B 321, 509 (1989).}
%
\Title{
\vbox{\baselineskip12pt\hbox{{\tt math-ph/0106005}}}}
{{\vbox {
\vskip-10mm
\centerline{The General $O(n)$ Quartic Matrix Model}
\vskip2pt
\centerline{and its application to Counting Tangles and Links}
}}}
\medskip
\centerline{P.~Zinn-Justin}\medskip
\centerline{\sl Laboratoire de Physique Th\'eorique et Mod\`eles Statistiques}
\centerline{\sl Universit\'e Paris-Sud, B\^atiment 100}
\centerline{\sl 91405 Orsay Cedex, France}
\vskip .2in
The counting of alternating tangles in terms of their crossing
number, number of external legs and connected components
is presented here in a unified framework using quantum field-theoretic methods
applied to a matrix model of colored links.
The overcounting related to topological equivalence of diagrams is
removed by means of a renormalization scheme of the matrix model;
the corresponding ``renormalization equations'' are derived.
Some particular cases are studied in detail and solved exactly.
\Date{06/2000}

\newsec{Introduction}
The goal of this paper is to investigate a fairly general enumeration
problem
related to the theory of knots, links and tangles:
we want to count objects which live in 3-dimensional space and are (loosely) made
of a certain collection of ``ropes'', some of which open (with fixed endpoints)
and some closed on themselves, intertwined together in an alternating way.
As usual in knot theory, these objects will be considered up
to topological equivalence (deformation or ambient isotopy), and
represented by their projections on the plane; we shall then
classify them according to the (minimal) number of crossings, the number
of connected components, and the way the various ``external legs'' connect
to each other, see for example Fig.~\typeex. 
\fig\typeex{A diagram with 3 open and 1 closed line,
intertwined together with 7 crossings.}{\epsfxsize=2cm\epsfbox{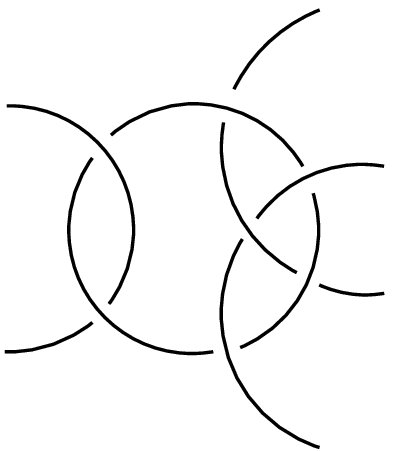}}
Without going into too much detail for now, we see that a convenient
way to keep track of the number of connected components and of the
connections of the external legs is to use colors, see Fig.~\typeextwo.
\fig\typeextwo{Coloring the diagram of Fig.~\typeex. Open lines
have fixed colors (distinct from each other), whereas
closed lines have arbitrary color.}{\epsfxsize=8cm\epsfbox{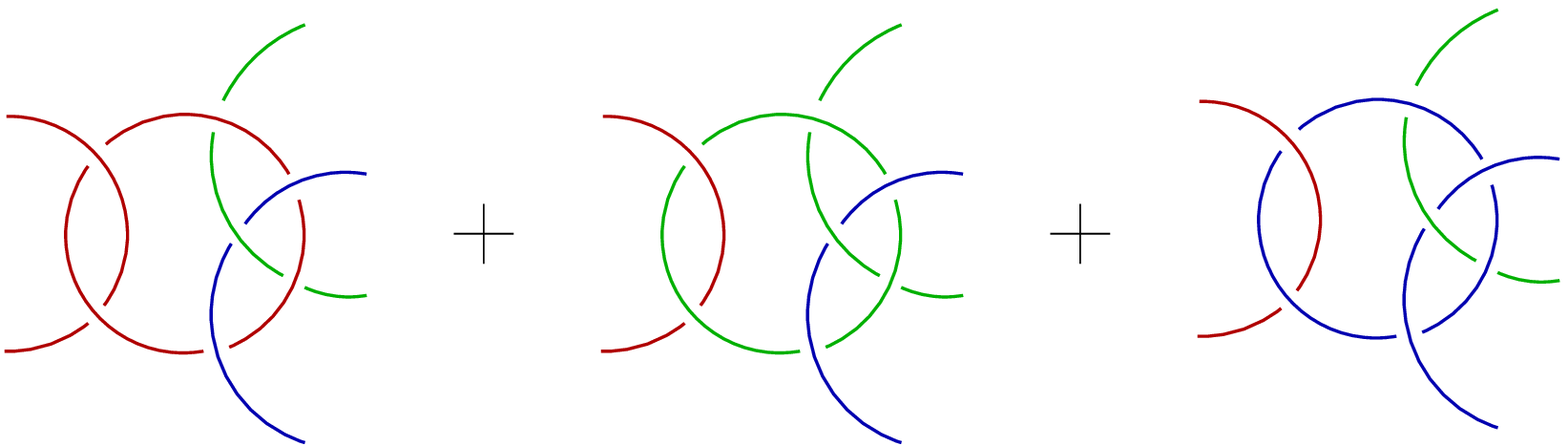}}
The colors allow us to distinguish the various external legs and add
an extra power series variable in the theory (the number of colors $n$) to count
separately objects with different numbers of connected components.

The idea to use colors was already suggested in \refs{\ZJZ,\PZJ} 
and present in the work
\ZJZb. At this stage it is natural to define a matrix model whose
Feynman diagram expansion will produce such diagrams with $n$ colors.
Here we shall give a unified quantum field-theoretic treatment
of this $O(n)$-invariant matrix model,
which simplifies and generalizes the equations
obtained in \ZJZb\ (section 2 below). In particular, it gives a practical way to
do the enumeration by computer;
this procedure was recently used in the numerical work \JZJ.

Even though the matrix model we propose is fairly natural, since as
we shall see it is the
most general quartic $O(n)$-invariant matrix model with a single trace in the
action, it is in general unsolvable (or at least unsolved).
It can be thought of as describing
a statistical model on random dynamical lattices; more precisely,
it is a model of fully packed loops drawn in $n$ colors on random tetravalent
planar diagrams with weights attached to vertices (intersections or tangencies
of loops). Even the corresponding model on a 
regular (flat) square lattice is not fully understood. 
However it is tempting to speculate
on its universality class; and that putting it on random lattices will correspond
to the usual coupling of two-dimensional conformal field theory to gravity,
which allows to predict the critical exponents of the theory
based on the KPZ relation \KPZ. This in turn leads
to various conjectures on the asymptotic number of large links and tangles,
made in \PZJ, which have been checked numerically in \JZJ.
We shall not come back to these conjectures here, but instead produce
exact analytic solutions of two particular cases of our matrix model
(section 3 below):
the classical case $n=1$ (no colors), with some generalizations of
the results of \ZJZ; and the case $n=-2$, which is interesting because
its asymptotic behavior cannot be obviously guessed by the universality
arguments mentioned above.

\newsec{General principle}
We assume the reader familiar with the concept of links and tangles. Let us
recall here that once projected on a plane, they give rise to 
{\it planar diagrams}\/
with tetravalent vertices which must be ``decorated'' to distinguish
under/over-crossings. Link diagrams are closed, whereas tangle diagrams
have external legs. The diagrams are said to be alternating if one meets
undercrossings and overcrossings alternatingly as one follows the various
closed loops of the diagram. The alternating property allows to ignore
the decorations of the vertices since they can be recovered from the diagram
alone (up to a mirror symmetry for the closed diagrams, see below).

\subsec{Definition of the $O(n)$ matrix model}
As in \PZJ\ and \ZJZb,
we start with the following matrix integral over $N \times N$ hermitean matrices 
\eqn\mmm{
Z^{(N)}(n,g)=\int\! \prod_{a=1}^n \d M_a
\, \E{N\,\tr\left(-{1\over 2} \sum_{a=1}^n M_a^2+{g\over 4} \sum_{a,b=1}^n
M_a M_b M_a M_b\right)}}
where $n$ is (for now) a positive integer. The integral is normalized so
that $Z^{(N)}(n,0)=1$.
The partition function \mmm\ displays a $O(n)$ symmetry
where the $M_a$ form a vector of $O(n)$.

Expanding in power series in $g$ generates
Feynman diagrams with double edges (``fat graphs'') drawn in $n$ colors
in such a way that colors cross each other at the vertices.
By taking the large $N$ limit one selects the planar diagrams,\foot{Note that
the next orders in the $1/N$ expansion of the free energy $\log Z^{(N)}$ would
correspond to link diagrams drawn on thickened surfaces of higher genus,
cf \KAUFb.}
which are closely related to alternating link diagrams, cf Fig.~\feya.
\fig\feya{A planar Feynman diagram of \mmm\ and the corresponding
alternating link diagram.}{\epsfxsize=8cm\epsfbox
{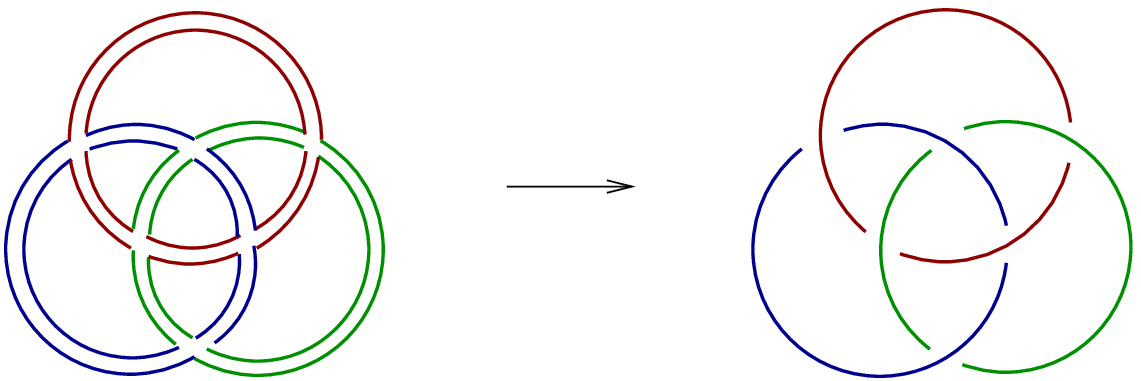}}
More precisely, the large $N$ ``free energy''
\eqn\mmmb{F(n,g)=\lim_{N\to\infty}{\log Z^{(N)}(n,g)\over N^2}}
is a double generating function of the number $f_{k;p}$ of alternating 
link diagrams with
$k$ connected components and $n$ crossings (weighted by the inverse of their
symmetry factor, and with mirror images identified):
\eqn\dblgen{F(n,g)=\sum_{k=1}^\infty\sum_{p=1}^\infty f_{k;p}\, n^k g^p}
Note that it is clearly
possible to analytically continue $Z^{(N)}(n,g)$ to arbitrary values of $n$
(using, for example, a \HS\ transformation) so that Eq.~\dblgen\ still holds.
In particular, the
counting of knot diagrams is given by $F_{1;p}$ and can be obtained by
formally taking the limit $n\to 0$, in the spirit of the replica
method. Also, if $n$ is an even negative integer one can write fermionic
analogues of \mmm, see section 3.2, which display $Sp(|n|)$ symmetry.

If one is interested in counting objects with a weight of $1$,
one cannot consider the free energy
which corresponds to closed diagrams, but instead correlation
functions of the model which generate diagrams with external
legs: these are essentially tangle diagrams.
Typically, we shall be interested in the two-point function
\eqn\two{G(n,g)\equiv \lim_{N\to\infty}\left<{1\over N} \tr M_a^2\right>
}
where the measure on the the matrices $M_a$ is given by Eq.~\mmm\ 
and $a$ is any fixed index, which generates tangle diagrams with two
external legs; and the connected
four-point functions
\eqna\four
$$\eqalignno{
\Gam_1(n,g)&=\lim_{N\to\infty}\left< {1\over N}\tr (M_a M_b)^2\right>&\four{.1}\cr
\Gam_2(n,g)&=\lim_{N\to\infty}\left< {1\over N}\tr (M_a^2 M_b^2)\right>-G(n,g)^2&\four{.2}\cr
}$$
where $a$ and $b$ are two distinct indices, which generate tangle diagrams with four
external legs of type $1$ and $2$ (see Fig.~\types). 
Note that the freedom to replace a link diagram with its mirror image by
inverting all under/over-crossings is, in the case of correlation functions,
removed by fixing conventionally the first
crossing encountered starting from a given external leg.

Let us briefly mention for now that the definition of $G(n,g)$ again
assumes $n$ to be a positive integer, and has a natural continuation to any $n$;
however the definitions of $\Gam_i(n,g)$ are only meaningful for $n$ integer
greater or equal to $2$, and there is a difficulty associated to this,
which will be explained in section 2.3.
\fig\types{Tangles of types 1 and 2.}{\epsfxsize=8cm\epsfbox{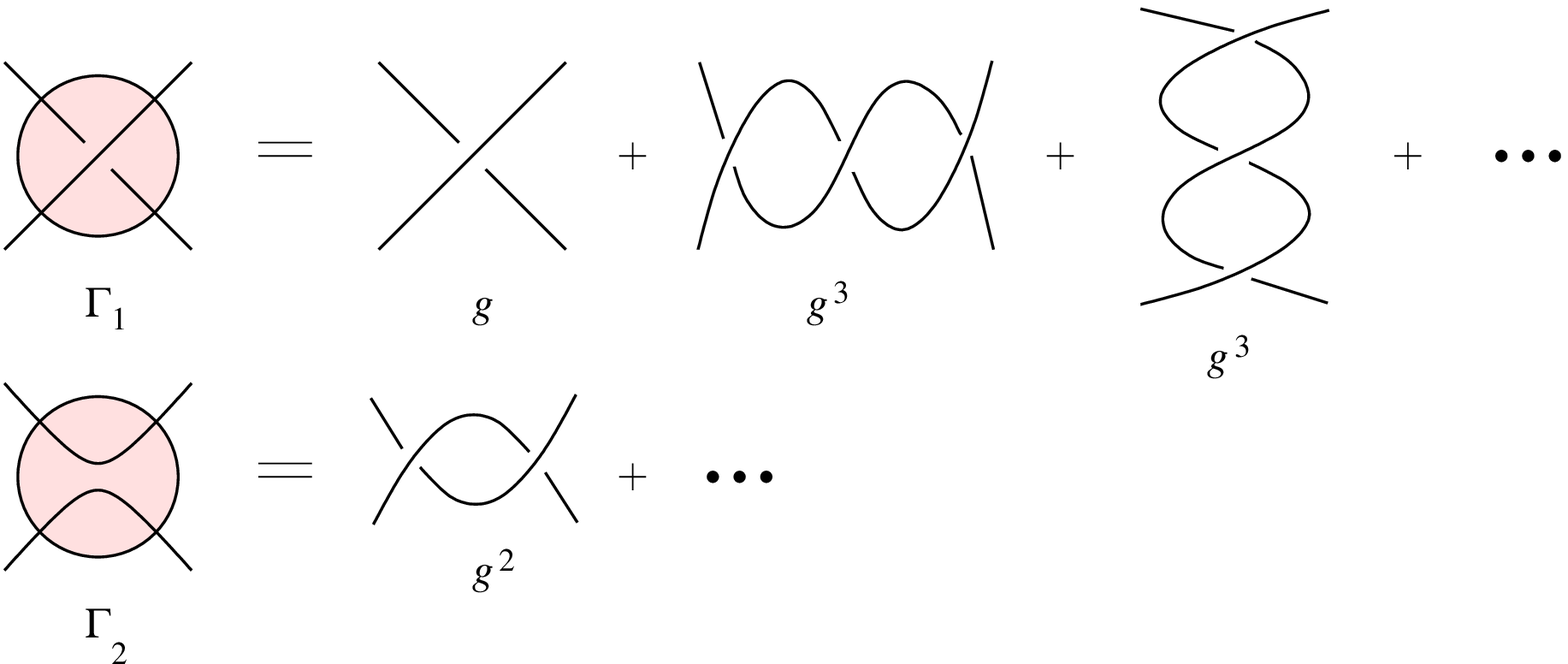}}

\subsec{Renormalization of the $O(n)$ model}
The model presented above is not sufficient to properly count colored
tangles.
Essentially, this comes from the fact that there is not a one-to-one
correspondence between diagrams and the objects they are obtained from
by projection. This generates a redundancy in the counting since
to a given knot will correspond many (an infinity of) diagrams, each
counted once.
In the case of alternating diagrams one can distinguish
two steps to remove this redundancy. First one must find a way to
restrict ourselves to {\it reduced}\/ diagrams which contain no irrelevant
crossings (Fig.~\nprime~a)); such diagrams will have minimum number of crossings.
It turns out to be convenient to introduce at this point a closely
related notion: a link is said to be {\it prime}\/ if it cannot be
decomposed into two pieces in the way depicted on Fig.~\nprime~b). It is clear
that at the level of diagrams, forbidding decompositions of the type
of Fig.~\nprime~b) automatically implies that the diagram is reduced; and
we shall therefore restrict ourselves to prime links and tangles.
\fig\nprime{a) An irrelevant crossing. b) A non-prime link.}{\epsfxsize=6cm\epsfbox{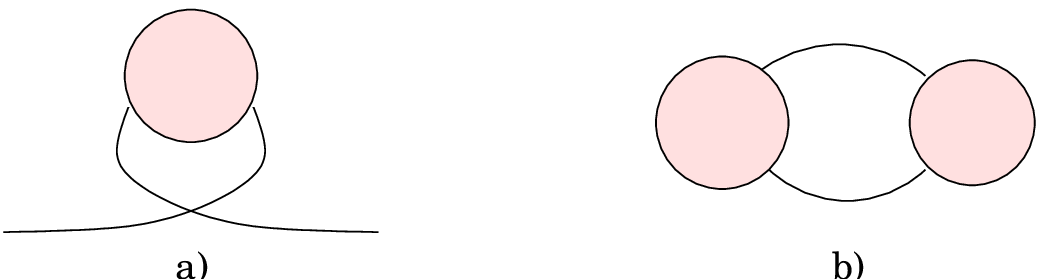}}
There may still be several reduced diagrams corresponding to the same
link: according to the flyping conjecture, proved in \MTh, two such diagrams
are related by a finite sequence of flypes, see Fig.~\flyp.
\fig\flyp{A flype.}{\epsfxsize=6cm\epsfbox{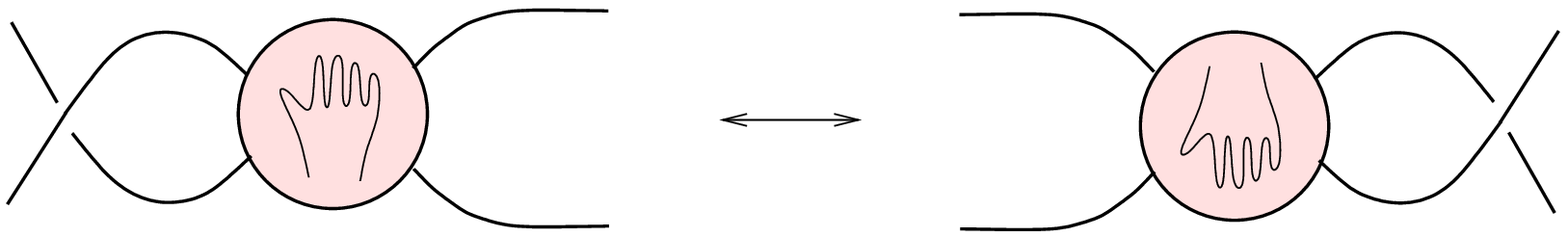}}

To summarize, there are two problems: a) the diagrams generated by applying
Feynman rules are not necessarily reduced or prime; b) several
reduced diagrams may correspond to the same knot due to the flyping
equivalence. A study of Figs.~\nprime\ and \flyp\ shows that this
``overcounting'' is local in the diagrams in the sense that problem
a) is related to the existence of sub-diagrams with 2 external legs,
whereas problem b) is related to a certain class of sub-diagrams with
4 external legs. Clearly such graphs can be cancelled by the inclusion
of appropriate {\it counterterms}\/ in the action.
We are therefore led to the conclusion that we must
{\it renormalize}\/ the quadratic and quartic interactions of
\mmm. Now renormalization theory tell us that we should include
in the action from the start every term compatible with the symmetries
of the model, since they will be generated dynamically by the 
renormalization. In order to preserve connectedness we only look for terms
of the form of a single trace.
A key observation is that, while there is only one such quadratic
$O(n)$-invariant term, there are {\it two}\/ quartic $O(n)$-invariant
terms, which leads to a generalized model with 3 coupling constants
in the action (bare coupling constants):
\eqn\mmmgen{
Z^{(N)}(n,t,g_1,g_2)=\int\! \prod_{a=1}^n \d M_a
\, \E{N\,\tr\left(-{t\over 2} \sum_{a=1}^n M_a^2
+{g_1\over 4} \sum_{a,b=1}^n (M_a M_b)^2
+{g_2\over 2} \sum_{a,b=1}^n M_a^2 M_b^2
\right)}}
The Feynman rules of this model now allow loops of different colors to ``avoid''
each other, which one can imagine as tangencies (Fig.~\feyren).
\fig\feyren{Vertices of the generalized $O(n)$ matrix model.}{%
\epsfxsize=4cm\epsfbox{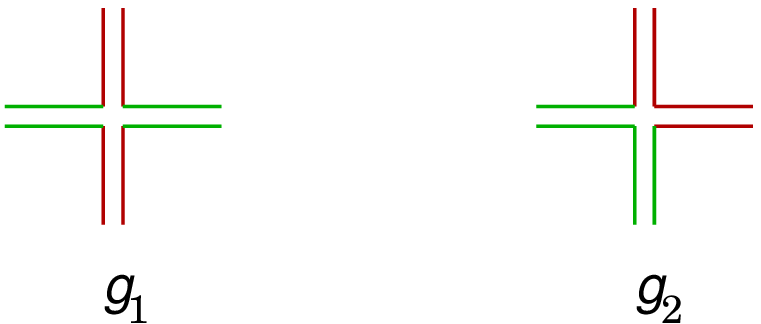}}
We define
again the correlation functions $G(n,t,g_1,g_2)$ and $\Gam_i(n,t,g_1,g_2)$
(Eqs.~\two\ and \four{}), and want to extract from them
the counting of colored alternating tangles with external legs.

The idea is to find the expressions of $t(g)$, $g_1(g)$ and
$g_2(g)$ as a function of the renormalized coupling constant $g$,
in such a way that the overcounting is suppressed and the correlation
functions are generating series in $g$ of the number of colored
tangles. At leading order, we shall
have $t(g)=1+o(1)$, $g_1(g)=g+o(g)$ and $g_2(g)=o(g)$ so that we recover
the original model \mmm. However there will be higher order corrections
which correspond to the counterterms.

Let us consider $t(g)$ first. It is clear that one must remove all two-legged
subdiagrams, that is impose
\eqn\ren{
G(n,t,g_1,g_2)=1
}
\fig\two{Decomposition of the two-point function. Reexpanding in powers
of $t-1$ will cancel the powers of $\Sigma$ iff $t=1+\Sigma$.}{\epsfxsize=10cm\epsfbox{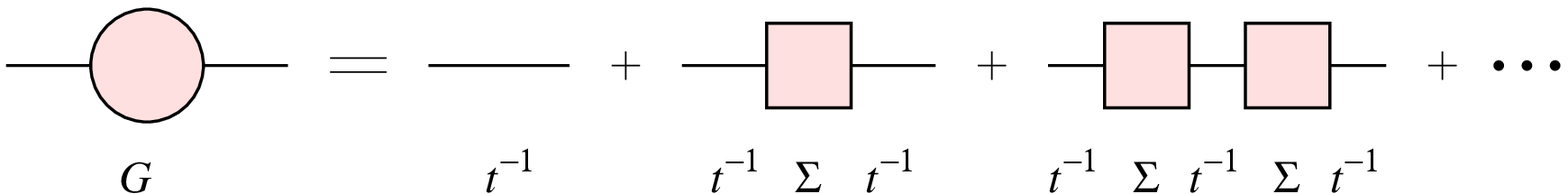}} 

Let us see more explicitly how this fixes $t(g)$.
Noting that (Fig.~\two)
\eqn\onepi{
G={1\over t-\Sigma}
}
where $\Sigma$ is the generating function of 1PI (one-particle irreducible,
i.e.\ which cannot be made disconnected by removing one edge) 
two-legged diagrams, one finds equivalently that
\eqn\renb{
t(g)=1+\Sigma(g)
}
i.e.\ the counterterms generated by $t(g)$ must cancel all 1PI two-legged
subdiagrams. This is almost a tautology; notice however that one must {\it not}\/
cancel all two-legged subdiagrams, since one-particle reducible
diagrams would be subtracted multiple times.

\fig\elemflyp{Breaking a flype into two elementary flypes.}{\epsfxsize=12cm\epsfbox{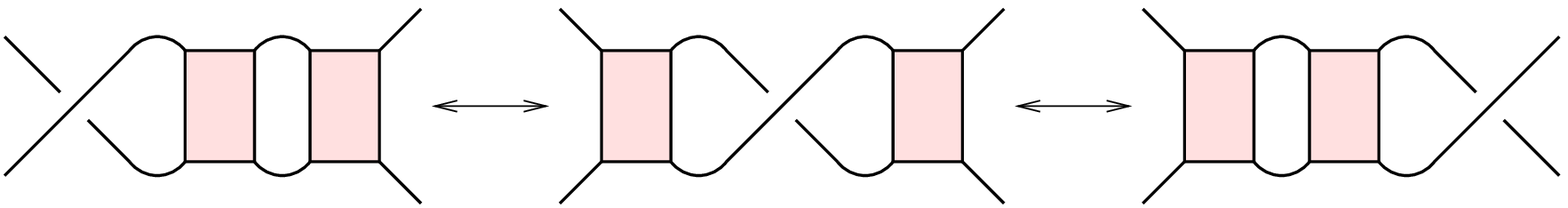}}
Next, we must consider the flyping equivalence. Again, it is important to
notice that a flype can be made of several ``elementary'' flypes (Fig.~\elemflyp), 
an elementary flype being by definition one that cannot be decomposed 
any more in
this way. In the terminology of QFT, an elementary flype 
consists precisely of
one simple vertex connected by two edges to a non-trivial H-2PI 
(two-particle irreducible in the horizontal channel) tangle diagram.
Non-trivial means not reduced to a single vertex; H-2PI means that
the tangle diagram cannot be cut
into two pieces containing the left and right external legs respectively,
by removing two edges.
We therefore need to introduce
auxiliary generating functions $H'_1(g)$, $H'_2(g)$ and $V'_2(g)$
for non-trivial H-2PI tangles of type 1, of type 2 and of type 2 rotated by
$\pi/2$ respectively. Only these must be included in the counterterms.
It is now a simple matter to consider all possible insertions of elementary
flypes as tangle sub-diagrams of a diagram; 
taking into account the two types of tangle sub-diagrams and
the two channels (horizontal and vertical), we find (Fig.~\renflyp)
that the renormalization of $g_1$ and $g_2$ is
simply:
\eqna\renc
$$\eqalignno{
g_1(g)&=g(1-2H'_2(g))&\renc{.1}\cr
g_2(g)&=-g(H'_1(g)+V'_2(g))&\renc{.2}\cr
}
$$
\fig\renflyp{Counterterms needed to cancel flypes.}{\epsfxsize=12cm\epsfbox{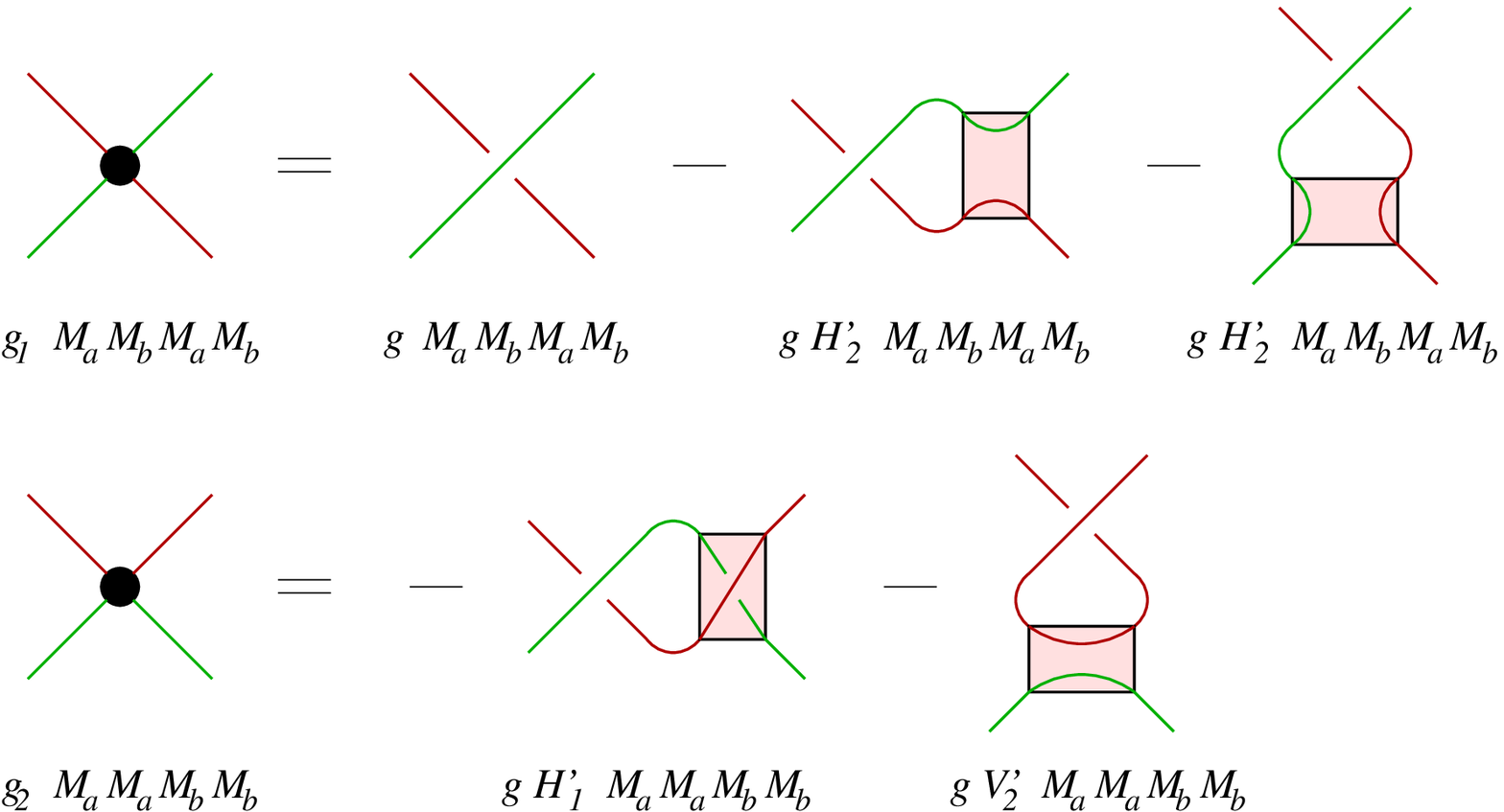}}
All that is left is to find the expressions of the auxiliary generating
functions in terms of known quantities. They are easily obtained
by decomposing the four-point functions in the horizontal and vertical channels,
and will not be rederived here (the reader is referred to e.g.\ \ZJZb\ for
details).
\eqna\deci
$$\eqalignno{
H'_2\pm H'_1 &=1-{1\over (1\mp g)(1+\Gam_2\pm\Gam_1)}
&\deci{\rm a}\cr
H'_2+nV'_2+H'_1&=1-{1\over (1-g)(1+(n+1)\Gam_2+\Gam_1)}
&\deci{\rm b}\cr
}$$

\subsec{Summary and discussion}
We shall now summarize and rewrite more explicitly the formulae found previously,
as well as discuss their implications.

Let us assume that for a certain $n$, we have computed the
free energy $F(n,t,g_1,g_2)$. What can we extract from the formulae
of the previous paragraph, and how?

First, let us differentiate $F$; we find
\eqn\diff{
G=-2{\der\over\der t} {F\over n}
}
as well as two other quantities,
\eqna\diffb
$$\eqalignno{
F_1&=4{\der\over\der g_1} {F\over n}
={1\over n}\lim_{N\to\infty}\left< {1\over N}\tr \sum_{a,b} (M_a M_b)^2\right>
&\diffb{.1}\cr
F_2&=2{\der\over\der g_2} {F\over n}
={1\over n}\lim_{N\to\infty}\left< {1\over N}\tr \sum_{a,b} M_a^2 M_b^2\right>
&\diffb{.2}\cr
}$$
According to the equations of motion, these three quantities are not independent:
\eqn\eom{
t G=1+ g_1 F_1+2g_2 F_2
}

Comparing \diffb{} with the definition \four{} of the $\Gam_i$, we
see that there are two different choices of basis of the
four-point functions;\foot{Using the $\Gam_i$ as the preferred
basis is not only natural diagramatically; it is also imposed
by the structure of the equations such as \renc{} and \deci{}.}
using $O(n)$-invariance of the measure it is
easy to relate them:
\eqna\diffc
$$\eqalignno{
F_1&=n\Gam_1+2(\Gam_2+G^2)&\diffc{.1}\cr
F_2&=\Gam_1+(n+1)(\Gam_2+G^2)&\diffc{.2}\cr
}$$
These relations also have a simple diagrammatic interpretation,
which proves in particular that they are valid for any (complex) $n$.
One observes that relations \diffc{} can be inverted to extract
$\Gam_1$ and $\Gam_2$ only if $n\ne 1,-2$. These two cases will
be the object of study of the next section, they are the first
in the series of bosonic / fermionic matrix models; and as will be shown
these are the values of $n$ for which the model possesses only one
quartic $O(n)$-invariant, contrary to the generic case. For now
we simply observe that for $n=1$, $F_1=F_2$ and therefore
according to \diffb{}, the free energy $F$ is a function of
$g_1+2g_2$ only; while for $n=-2$, $F_1=-2F_1$ and $F$ is a function of
$g_1-g_2$ only.

Once we have computed $G$, $\Gam_1$ and $\Gam_2$,
we can slightly simplify the renormalization equations using
the obvious scaling property:
\eqna\scal
$$\eqalignno{
G(n,t,g_1,g_2)&={1\over t} G(n,1,g_1/t^2,g_2/t^2)&\scal{a}\cr
\Gam_i(n,t,g_1,g_2)&={1\over t^2} \Gam_i(n,1,g_1/t^2,g_2/t^2)&\scal{b}\cr
}$$
Combining this with Eq.~\ren\ results in fixing $t(g)$:
\eqn\scalb{
t(g)=G(n,1,g_1(g)/t(g)^2,g_2(g)/t(g)^2)
}
At this stage the three unknowns $t(g)$, $g_1(g)$, $g_2(g)$ only
appear through the combinations $h_1(g)\equiv g_1(g)/t(g)^2$, 
$h_2(g)\equiv g_2(g)/t(g)^2$; in particular we have
the following expressions for the $\Gam_i\equiv\Gam_i(n,t(g),g_1(g),g_2(g))$:
\eqn\scalc{
\Gam_i={\Gam_i(n,1,h_1(g),h_2(g))\over
G(n,1,h_1(g),h_2(g))^2}
}
We only need to solve the two remaining renormalization equations
\renc{}, which we rewrite here:
\eqna\rencc
$$\eqalignno{
h_1(g)\, G(n,1,h_1(g),h_2(g))^2&=g(1-2H'_2(g))&\rencc{.1}\cr
h_2(g)\, G(n,1,h_1(g),h_2(g))^2&=-g(H'_1(g)+V'_2(g))&\rencc{.2}\cr
}
$$
where the auxiliary generating functions are still given in terms of the
$\Gam_i$ by Eqs.~\deci{}.

Finally, solving Eqs.~\rencc{} gives access to the $\Gam_i$,
which are the generating series of the numbers of prime alternating tangles
of type $i$.
However, we can go further.
By computing other correlation functions in the model and composing
them with the solutions $t(g)$, $g_1(g)$, $g_2(g)$ of the equations above,
one can extract
the generating functions of the number of alternating tangles with an
arbitrary number of external legs. The correlation functions we
consider are traces of non-commutative words in the $M_a$ of degree $2k$
(for $2k$ external legs). We usually restrict ourselves to {\it connected}\/
correlation functions (free cumulants in the language
of free probabilities), which exclude configurations in which some strings have
no crossings with the other strings and can be pulled out altogether. This
choice is only a matter of taste.

For example, there are five $O(n)$-invariants
of degree 6, except, as before, for a finite set of values of $n$
for which there are fewer: only $4$ for $n=-4$, $3$ for $n=2$, $2$ for $n=-2$, 
$1$ for $n=1$. They are given by:
\eqna\typesixfor
$$\eqalignno{
\Xi_1&=\lim_{N\to\infty}\left<{1\over N}\tr(M_a M_b M_c M_a M_b M_c)\right>-{\rm disc.\ terms}&
\typesixfor{.1}\cr
\Xi_2&=\lim_{N\to\infty}\left<{1\over N}\tr(M_a M_b M_c M_a M_c M_b)\right>-{\rm disc.\ terms}&
\typesixfor{.2}\cr
\Xi_3&=\lim_{N\to\infty}\left<{1\over N}\tr(M_a M_a M_b M_c M_b M_c)\right>-{\rm disc.\ terms}&
\typesixfor{.3}\cr
\Xi_4&=\lim_{N\to\infty}\left<{1\over N}\tr(M_a M_b M_b M_a M_c M_c)\right>-{\rm disc.\ terms}&
\typesixfor{.4}\cr
\Xi_5&=\lim_{N\to\infty}\left<{1\over N}\tr(M_a M_a M_b M_b M_c M_c)\right>-{\rm disc.\ terms}&
\typesixfor{.5}\cr
}$$
($a$, $b$, $c$ distinct) and give rise to the various six-legged diagrams depicted
on Fig.~\typesix.
\fig\typesix{The five types of tangles with 6 external legs.}
{\epsfxsize=8cm\epsfbox{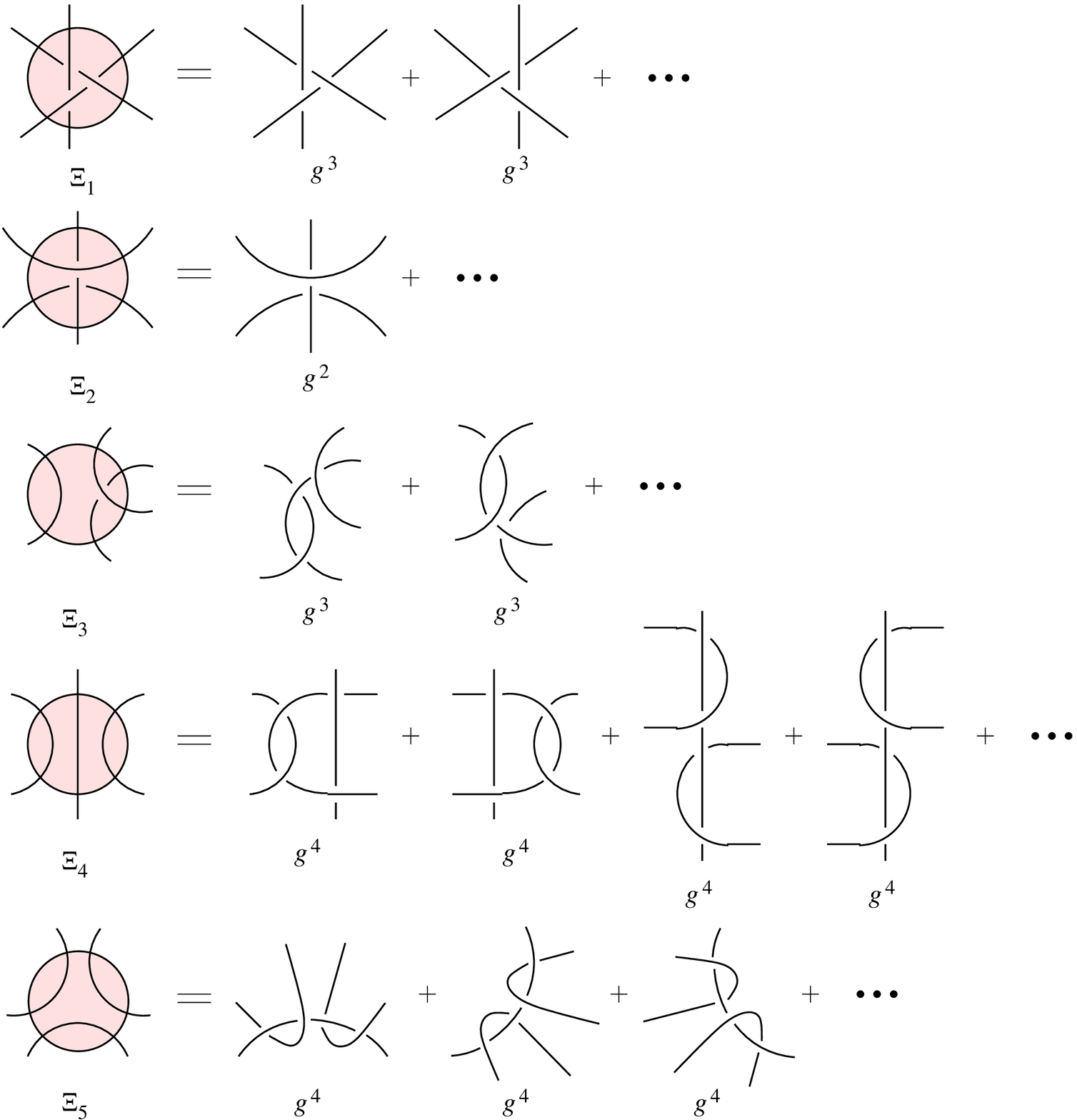}}

\newsec{Application: two solvable cases}
There are currently two values of $n$ for which the corresponding
matrix model has been exactly solved: $n=1$ and $n=2$. The case $n=1$
is particularly important since it corresponds to counting all alternating
tangles regardless of the number of connected components; we shall
investigate it here in detail, generalizing known results \refs{\STh,\ZJZ}.

The application of the $O(n=2)$ matrix model (also known as six-vertex
model on dynamical random lattices) to knot theory has already been made
in \ZJZb, using slightly different methods than in the present paper,
and we shall not come back to it.

However, we have found earlier that aside from $n=1$, there is another
special value of $n$, namely $-2$,
for which a simplification in the model occurs and we can expect
some exact analytic results.
We shall present below an analysis of this case.

\subsec{The case $n=1$: the usual tangles, and more}
As an illustration of the general principle developed above,
we present an elementary solution of the case $n=1$,
that is the counting of alternating tangles. Since there are
no colors one cannot distinguish the way the various external legs are connected;
the correlation functions available to us will be specified by the number of
external legs only.

Note that this solution, which generalizes the original calculation
of the number of prime alternating tangles with 4 external legs found in \STh,
is technically different from it.

We start by setting $n=1$ in the definition of the partition function
(Eq.~\mmmgen); we find:
\eqn\mmone{
Z^{(N)}(t,g_0)=\int \d M
\, \E{N\,\tr\left(-{t\over 2} M^2
+{g_0\over 4} M^4 \right)}}
where $g_0\equiv g_1+2g_2$. The fact that the partition function only depends
on a particular combination of $g_1$ and $g_2$ is consistent with what was found
in Section 2.3 and related to the existence of only one quartic
$O(n)$-invariant for $n=1$.
The most general ``planar'' correlation functions of the model are of the form
\eqn\corone{
G_{2\ell}(t,g_0)\equiv \lim_{N\to\infty}\left<{1\over N} \tr M^{2\ell}\right>
}
for which we introduce the generating function:
\eqn\resone{
\omega(\lambda)\equiv \lim_{N\to\infty}\left<{1\over N} \tr {1\over\lambda-M}\right>
={1\over \lambda}+\sum_{\ell=1}^\infty G_{2\ell}{1\over\lambda^{2\ell+1}}
}
and the corresponding connected correlation functions $G_{2\ell}^c$, whose
generating function is the inverse function $\lambda(\omega)$:
\eqn\rescone{
\lambda(\omega)={1\over\omega}+\sum_{\ell=1}^\infty G_{2\ell}^c \omega^{2\ell-1}
}
Among them we have the two-point function $G\equiv G_2^c=G_2$ and the
connected four-point function $\Gam\equiv G_4^c=G_4-2G_2^2$
which is nothing
but the generating function of all tangles (regardless of type):
$\Gam=\Gam_1+2\Gam_2$.
Since we do not
have access to $\Gam_1$ and $\Gam_2$ separately, we need to recombine the
equations of Section 2.2 so that only $\Gam$ appears in them.
Fortunately, this turns out to be possible; 
taking \renc{.1}$+2\times$\renc{.2} results in
\eqn\renone{
g_0(g)=g(1-2H(g))
}
where $H(g)\equiv H'_2(g)+H'_1(g)+V'_2(g)$ is the generating
function of all H-2PI non-trivial tangles. Eq.~\renone\ can of course be
derived directly in a manner similar to Eqs.~\renc{}, by simply disregarding
the types of the tangles i.e.\ how the outgoing strings are connected to
each other inside the tangle.

Setting $n=1$ in Eq.~\deci{b}, we also find that
\eqn\renoneb{
H(g)=1-{1\over(1-g)(1+\Gam)}
}
so that for $n=1$ (and $n=1$ only) we have a closed subset of equations.

We now turn to the solution of our matrix model.
We do not repeat the calculation of the $G_{2\ell}$ here since it is a standard
result of matrix models, see \BIPZ.
Starting from the following expression:
\eqn\resoneb{
\omega(\lambda)={1\over2}\left(t\lambda-g_0\lambda^3
-(t-g_0\lambda^2-g_0 A)\sqrt{\lambda^2-2A}\right)
}
with $A={1\over3}{t-\sqrt{t^2-12g_0}\over g_0}$ solution of
\eqn\eqA{
3A^2 g_0-2At+4=0
}
we find that
\eqn\coroneb{
G_{2\ell}=A^\ell {(2\ell-1)!!\over (\ell+2)!}\left(2(\ell+1)-{\ell\over2} A t\right)
}
In particular $G={1\over6}A(4-{At\over2})$, and since $G=1$ according
to Eq.~\ren, we can express $t$ as a function of $A$:
\eqn\eqAb{
t={4\over A^2}(2A-3)
}

Similarly, using the explicit expression of $\Gam=G_4-2$, plugging it into 
Eqs.~\renone, \renoneb,
and using Eqs.~\eqA, \eqAb\ to express $t$ and $g_0$ in terms of $A$
results in the following fifth degree equation for $A$:
\eqn\eqAc{
32(1-g)-64(1-g)A+32(1-g)A^2-4(1+2g-g^2)A^3+6g(1-g)A^4-g(1-g)A^5=0
}
$A(g)$, specified by Eq.~\eqAc\ and $A(g=0)=2$,
is a well-defined analytic function of $g$ in a neighborhood of $g=0$.
The data of $A(g)$ is enough to recover all correlation functions since
we have, combining Eqs.~\coroneb\ and \eqAb:
\eqn\coronec{
G_{2\ell}=2 A^{\ell-1} {(2\ell-1)!!\over(\ell+2)!}\left(3\ell-(\ell-1)A\right)
}
Similarly, one can extract the connected correlation functions,
using the fact that $\lambda(\omega)$ satisfies a cubic equation
(cf Eq.~\resoneb); after a tedious calculation one finds
\eqn\coroned{
G_{2\ell}^c= {c_\ell\over \ell!} (A-2)^{\ell-1} \left(3\ell-2-(\ell-1)A\right)
}
where $c_\ell$ is a constant (which already appeared in \BIPZ):
\eqn\const{
c_{\ell+1}={1\over 3\ell+1}\sum_{\ell/2\le q\le \ell} (-4)^{q-\ell} 
{(\ell+q)!\over (2q-\ell)!(\ell-q)!}
}
This concludes the calculation of the generating series of the number
of tangles with any given number of external legs. In the appendix,
the first few orders of $G_4^c$, $G_6^c$, $G_8^c$ are given.

Let us now discuss the asymptotic behavior of the coefficients of the various
series for which we found an exact expression. All are simple polynomials in
$A(g)$, so that we need to study the latter only. As can be easily checked,
the singularity of $A(g)$ closest to the origin is the usual singularity
of 2D pure gravity, which is given by $g_{0c}=4/27$ and $t_c=4/3$,
so that $A_c=3$, and, plugging into Eq.~\eqAc,
\eqn\sing{
g_c={\sqrt{21001}-101\over 270}
}
We expand $A$ around $g\uparrow g_c$ and find
\eqn\singexp{
A=3-a\,(g_c-g)^{1/2}+b\,(g_c-g)+O((g_c-g)^{3/2})
}
with ($a>0$)
\eqna\expconst
$$\eqalignno{
a^2&={2877137 + 7087 \sqrt{21001}\over339696}&\expconst{a}\cr
b&= { 5 (99397733 + 2127733 \sqrt{21001})\over 901510722}\cr
}$$
This provides the leading singular part of $G_{2\ell}^c$:
\eqn\singpart{
G_{2\ell}^c={\rm reg}+ {c_\ell\over(\ell-2)!} a \Big(b+{1\over3}(\ell-2)a^2\Big)
(g_c-g)^{3/2}+\cdots
}
which finally yields the large order behavior of $G_{2\ell}^c$:
if $G_{2\ell}^c=\sum_{p=1}^\infty \gamma_{2\ell;p} g^p$ then
\eqn\asy{
\gamma_{2\ell;p}{\buildrel p\to\infty\over\sim}
 {3\over4\sqrt{\pi}}
{c_\ell\over(\ell-2)!} a \Big(b+{1\over3}(\ell-2)a^2\Big)
p^{-5/2} g_c^{3/2-p}
}
For $\ell=2$ this result coincides with the theorem 1 of \STh.
Note that for any $\ell$ the asymptotic behavior
is the same up to a constant. One can of course send $\ell$ and $p$ to infinity
in a correlated manner to obtain a non-trivial scaling limit (here,
$\ell\propto p^{1/2}$); but the result
is known to be universal and
is therefore the usual scaling loop function of pure gravity,
which will not be reproduced here.

\subsec{The case $n=-2$: a fermionic matrix model}
For $n$ negative
even integer, it is natural, in the spirit of supersymmetry,
to look for realizations of our model of colored links
under the form of a {\it fermionic}\/ matrix model with $Sp(-n)$ symmetry.
Let us show how this works in the simplest case, that is $n=-2$.

Our fields will be a ``complex fermionic matrix'', that is a matrix
$\Psi=(\Psi_{ij})$ where the $\Psi_{ij}$ are independent Grassmann
variables, and its formal adjoint $\Psi^\dagger=(\bar{\Psi}_{ji})$, which
together form the fundamental representation of $Sp(2)$. We apply to them
the usual rules of Berezin integration. We must next look
for $Sp(2)$-invariant quadratic and quartic invariants of the form
$\tr P(\Psi,\Psi^\dagger)$. Since the $\Psi$ are non-commutative objects,
one must consider arbitrary tensor products of the (dual of the) fundamental 
representation; however the trace property combined with the 
anti-commutativity of the matrix elements implies that 
an elementary circular permutation must have eigenvalue $-1$ in this
representation. Very explicitly, there are one
quadratic invariant, $\Psi\Psi^\dagger-\Psi^\dagger\Psi$, and two
independant quartic invariants, say 
$\Psi\Psi^\dagger\Psi\Psi^\dagger
-\Psi^\dagger\Psi\Psi\Psi^\dagger
-\Psi\Psi^\dagger\Psi^\dagger\Psi
+\Psi^\dagger\Psi\Psi^\dagger\Psi$ and
$\Psi\Psi\Psi^\dagger\Psi^\dagger
-\Psi^\dagger\Psi\Psi\Psi^\dagger
-\Psi\Psi^\dagger\Psi^\dagger\Psi
+\Psi^\dagger\Psi^\dagger\Psi\Psi$. 
However it is clear that the first quartic invariant
is stable by circular permuation and therefore its trace is zero. We are
thus left with the following expression:
\eqn\mmntwo{
Z^{(N)}(t,g_0)=\int \d\Psi\d\Psi^\dagger
\, \E{N\,\tr\left(-t \Psi\Psi^\dagger
+g_0 \Psi\Psi^\dagger\Psi^\dagger\Psi\right)}}
It is no surprise that the partition function only depends on one coupling
constant $g_0$, since the analysis of section 2.3 has shown us that
for $n=-2$ all large $N$ quantities depend only on the combination
$g_1-g_2$; and indeed,
by direct inspection one can identify $g_0=g_1-g_2$.

As in the case $n=1$, we have access to only one four-point fonction
\eqn\fourntwo{
\Gam\equiv\lim_{N\to\infty}
\left<{1\over N}\tr \Psi\Psi^\dagger\Psi^\dagger\Psi\right>_c
=\Gam_1-\Gam_2}
Again, a ``miracle'' happens in that a particular subset of the renormalization
equations becomes closed for $n=-2$; namely, taking \renc{.1}$-$\renc{.2}
\eqn\renntwo{
g_0(g)=g(1-2H'_2(g)+H'_1(g)+V'_2(g))
}
a combination of the H-2PI diagrams appears, which can be related to
$\Gam$ via the use of Eqs.~\deci{}:
\eqn\renntwob{
V'_2-2H'_2+H'_1=-2+ {3\over2(1+g)(1-\Gam)}
+{1\over2(1-g)(1+\Gam)}
}

We now briefly describe how to compute integral \mmntwo\ in the large $N$ limit.
We can set $t=1$ without loss of generality, as explained in section 2.3.
We perform the standard \HS\ transformation by introducing a hermitean matrix $A$:
\eqn\solntwo{
Z^{(N)}(t,g_0)=\int \d\Psi\d\Psi^\dagger \int \d A
\, \E{N\,\tr\left(- \Psi\Psi^\dagger -{1\over2} A^2
+\sqrt{-g_0}A(\Psi\Psi^\dagger+\Psi^\dagger\Psi)
\right)}}
The gaussian integral over $\Psi$ and $\Psi^\dagger$ can then be performed:
\eqn\solntwob{
Z^{(N)}(t,g_0)=\int \d A
\, \det(1\otimes 1+\sqrt{-g_0}(A\otimes 1+1\otimes A))\,\E{-{N\over2}\,\tr A^2}
}
We recognize at this point the usual $O(-2)$ 
fully packed (non-intersecting) loops model\foot{Of course
this might have been expected from the start,
since for any $n$ the $O(n)$ model of fully packed non-intersecting loops
corresponds to the particular case $g_1=0$ 
of our model, and therefore here $g_0=-g_2$.} \KoS.
We perform the change of variables: $M=(A-a_0)^2$ with $a_0=-{1\over 2\sqrt{-g_0}}$,
which absorbs the determinant, resulting in
\eqn\solntwoc{
Z^{(N)}(t,g_0)=\int \d M
\, \E{-{N\over2}\,\tr (\sqrt{M}+a_0)^2}
}
One can then diagonalize $M$ and compute the integral over eigenvalues
using standard large $N$ saddle point techniques. The resolvent of $M$ is
given by a complete elliptic integral of the third kind; in particular,
\eqn\solntwod{
G-{1\over 4g_0}+2=
\lim_{N\to\infty}\left<{1\over N} \tr M\right>
={1\over 96 \pi^4 g_0^2} K^3((8-8k^2+3k^4)K+4(k^2-2)E)}
where $K$ and $E$ are complete elliptic integrals of the first and
second kind with modulus $k$, and the coupling constant $g_0$ is given by
\eqn\solntwoe{
g_0=-{1\over 8\pi^2} K((2-k^2)K-2E)
}
(cf also \EK\ for a similar solution).
Finally, inserting the expression of $\Gam={1-G\over 2 g_0 G^2}+1$ obtained
from \solntwod\ into renormalization Eqs.~\renntwo--\renntwob\ results
in a $g$-dependent transcendental equation for the modulus $k^2$.
This equation is too complicated to be solved exactly; however it can
be easily solved numerically order by order.
Finally, $\Gam$
is the desired generating function $\Gam_1-\Gam_2$ of tangles;
in the appendix we present the first few orders of the expansion of $\Gam(g)$.

We now turn to the asymptotic behavior of the coefficients of $\Gam(g)$.
It is known that the $O(-2)$ matrix model does not have
any critical point of the usual form of 2D quantum gravity;
for example in \EK, the solution
of the $O(-2)$ model of dense loops, equivalent to ours,
is studied in detail when the elliptic modulus $k$ is in
the range $[0,1]$, which corresponds to $g_0\in [-\infty,0]$ in our 
notations, and no singularity is found. This does not mean, of course,
that $\Gam(g)$ has no singularity, since $g_0$ (and $g$) can move in the
whole complex plane. Generically, these singularities
are square root type singularities and given by
the equation ${\d g\over\d\Gam}=0$. It turns out that this equation has
plenty of solutions, though one can unfortunately
not write them down analytically. Numerically, one finds that the
the solutions with smallest modulus of $g$ are:
\eqn\critntwo{
g_c\approx -0.239 \pm 0.135 i
}
They are pairs of complex conjugate solutions: this indicates oscillatory
behavior of the series, which is due to the fact that $n<0$.
We therefore find that if $\Gam=\sum_{p=1}^\infty \gamma_p g^p$,
\eqn\critntwob{
\gamma_p\sim {\rm Re}({\rm cst}\, p^{-3/2} g_c^{-p})
}
where ${\rm cst}\approx -0.237\pm 0.090i $.
It would be interesting to find a physical interpretation of 
this critical point \`a la Yang--Lee.

Finally, let us note that one could combine the results of $n=+2$ \ZJZb\ and 
of $n=-2$:
this would give rise to a model of oriented tangles in which one counts
separately tangles with odd and even numbers of connected components.
Since the coefficients of $\Gam(g)$ (and presumably also of $\Gam_1(g)$
and $\Gam_2(g)$ separately) in the case $n=-2$ satisfy, according
to Eq.~\critntwob, $\gamma_p=O(3.64\ldots^p)$ 
whereas in the case $n=+2$ these coefficients
are of the order $6.28\ldots^p$, we conclude that there are, up to exponentially
small corrections, as many odd tangles as there are even tangles$\ldots$

\appendix{A}{Tables for $n=1$ and $n=-2$ up to $32$ crossings.}
\tab\resknot{Table of the number of prime alternating
tangles ($n=1$) with 4, 6, 8 external legs.}{\vbox{\offinterlineskip
\halign{\strut#&\enskip\vrule#\enskip%
&&\hfil{#\ }\hfil\crcr
$p$&&$G_c^4$&$G_c^6$&$G_c^8$\cr
\omit&height2pt\cr
\noalign{\hrule}
\omit&height2pt\cr
1& & 1& & \cr
 2& & 2& 3& \cr
 3& & 4& 14& 12\cr
 4& & 10& 
    51& 90\cr
 5& & 29& 186& 468\cr
 6& & 98& 708& 2196\cr
 7& & 372&
     2850& 10044\cr
 8& & 1538& 12099& 46170\cr
 9& & 6755& 53756& 
    215832\cr
 10& & 30996& 247911& 1029564\cr
 11& & 146982& 1178352& 
    5010192\cr
 12& & 715120& 5740224& 24830640\cr
 13& & 3552254& 
    28535604& 125073288\cr
 14& & 17951322& 144283404& 639037476\cr
 15& 
    & 92045058& 740126242& 3306068412\cr
 16& & 477882876& 3843972303& 
    17292904722\cr
 17& & 2508122859& 20180815236& 91335814848\cr
 18& &
     13289437362& 106957362161& 486589812240\cr
 19& & 71010166670& 
    571643594646& 2612379495996\cr
 20& & 382291606570& 3078146310603& 
    14122834373034\cr
 21& & 2072025828101& 16686687494650& 
    76829648302716\cr
 22& & 11298920776704& 91009054240656& 
    420345016423632\cr
 23& & 61954857579594& 499101633250932& 
    2311716994208856\cr
 24& & 341427364138880& 2750883342029780& 
    12773922263423472\cr
 25& & 1890257328958788& 15231756014050908& 
    70893591427443456\cr
 26& & 10509472317890690& 84695579659496748& 
    395034141129257304\cr
 27& & 58659056351295672& 472782954018549456& 
    2209407034450182552\cr
 28& & 328591560659948828& 2648662349568626736& 
    12399753592080373248\cr
 29& & 1846850410940949702& 
    14888203427107319436& 69813861782757325992\cr
 30& & 
    10412612510292744992& 83947527137925001240& 394245960540417041532\cr
 31& 
    & 58877494436409193754& 474714688448707647894& 
    2232568414958638372020\cr
 32& & 333824674188182988872& 
    2691749836124970938595& 12675855143073018219570\cr
}}}
\tab\resknotb{Table of the coefficients of $\Gam$ ($n=-2$ tangles with 
4 external legs).}{\vbox{\offinterlineskip
\halign{\strut#&\enskip\vrule#\enskip%
&&\hfil{#\ }\hfil\crcr
$p$&&$\Gam$\cr
\omit&height2pt\cr
\noalign{\hrule}
\omit&height2pt\cr
1& & 1\cr 2& & -1\cr 3& & 1\cr 4& & 1\cr 5& & -7\cr 6& 
    & 23\cr 7& & -51\cr 8& & 50\cr 9& & 212\cr 10& 
    & -1596\cr 11& & 6492\cr 12& & -19124\cr 13& & 
    37094\cr 14& & 1878\cr 15& & -437322\cr 16& & 2557800\cr 17& 
    & -10055712\cr 18& & 29767944\cr 19& & -58631365\cr 20& 
    & -4689017\cr 21& & 740682974\cr 22& & -4462194156\cr 23& 
    & 18243692937\cr 24& & -57186253699\cr 25& & 
    127394803329\cr 26& & -81353773012\cr 27& & -1062951245376\cr 28& 
    & 7538741871041\cr 29& & -33359417764221\cr 30& & 
    112902256367630\cr 31& & -286176860146756\cr 32& & 
    379259745656069\cr 
}}}
\footatend\vfill\supereject\immediate\closeout\rfile\writestoppt
\baselineskip=14pt\centerline{{\bf References}}\bigskip{\frenchspacing%
\parindent=20pt\escapechar=` \input refs.tmp\vfill\eject}\nonfrenchspacing
\bye